\documentclass[usenatbib]{mn2e}
\usepackage{epsfig,lscape,amsmath,color}

\newcommand{\etal}{et~al.} 

\newcommand{\UCHII}{UCH{\sc ii} }

\newcommand{\kms}{$\mbox{km~s}^{-1}$ }
\newcommand{\kmsns}{$\mbox{km~s}^{-1}$}

\newcommand{\Msol}{M$_{\odot}$}

\newcommand{\vsfig}[2]           
{
  \begin{center}
    \begin{minipage}[t]{0.05\textwidth}
      {\footnotesize \raisebox{40mm}{(#2)}}
    \end{minipage}
    \begin{minipage}[t]{0.42\textwidth}
      \psfig{file=./#1.ps,height=0.95\textwidth,angle=0}
    \end{minipage}
    \hfill
  \end{center}
}

\newcommand{\specdfig}[2]        
{
   \begin{center}
     \begin{minipage}[t]{0.45\textwidth}
         \psfig{file=eps/#1.eps,height=0.65\textwidth,width=1\textwidth,angle=0}
     \end{minipage}
     \hfill
     \begin{minipage}[t]{0.45\textwidth}
         \psfig{file=eps/#2.eps,height=0.65\textwidth,width=1\textwidth,angle=0}
     \end{minipage}
   \end{center}
}

\newcommand{\specsfig}[1]        
{
   \begin{center}
     \begin{minipage}[t]{0.45\textwidth}
         \psfig{file=eps/#1.eps,height=0.65\textwidth,width=1\textwidth,angle=0}
     \end{minipage}
   \end{center}
}

\newcommand{\boxfig}[1]        
{
   \begin{center}
     \begin{minipage}[t]{0.46\textwidth}
         \psfig{file=eps/#1.eps,height=0.45\textwidth,angle=0}
     \end{minipage}
   \end{center}
}

\newcommand{\twofig}[2]        
{
   \begin{center}
     \begin{minipage}[t]{0.5\textwidth}
         \psfig{file=eps/#1.eps,height=0.95\textwidth}
     \end{minipage}
     \hfill
     \begin{minipage}[t]{0.5\textwidth}
         \psfig{file=eps/#2.eps,height=0.95\textwidth}
     \end{minipage}
   \end{center}
}

\begin{document}

\title[6.7-GHz masers associated with star formation]{Confirmation of the exclusive association between 6.7-GHz methanol masers and high-mass star formation regions}
\author[S.\ L.\ Breen \etal]{S.\ L. Breen,$^{1}$\thanks{Email: Shari.Breen@csiro.au} S.\ P. Ellingsen,$^2$ Y. Contreras,$^1$ J.\ A. Green,$^1$ J.\ L. Caswell,$^1$ \newauthor J.\ B. Stevens,$^3$ J.\ R. Dawson,$^2$ M.\ A. Voronkov$^1$  \\
  \\
  $^1$ CSIRO Astronomy and Space Science, Australia Telescope National Facility, PO Box 76, Epping, NSW 1710, Australia\\
  $^2$ School of Mathematics and Physics, University of Tasmania, Private Bag 37, Hobart, Tasmania 7001, Australia\\
  $^3$ CSIRO Astronomy and Space Science, Australia Telescope National Facility, Locked Bag 194, Narrabri, NSW 239,0 Australia\\
}
 
 \maketitle
  
 \begin{abstract}

Recently, a comparison between the locations of 6.7-GHz methanol masers and dust continuum emission has renewed speculation that these masers can be associated with evolved stars. The implication of such a scenario would be profound, especially for the interpretation of large surveys for 6.7-GHz masers, individual studies where high-mass star formation has been inferred from the presence of 6.7-GHz methanol masers, and for the pumping mechanisms of these masers. We have investigated the two instances where 6.7-GHz methanol masers have been explicitly suggested to be associated with evolved stars, and we find the first to be associated with a standard high-mass star formation region, and the second to be a spurious detection. We also find no evidence to suggest that the methanol maser action can be supported in the environments of evolved stars. We thereby confirm their exclusive association with high-mass star formation regions.

\end{abstract}


\section{Introduction}

Since their initial discovery, methanol masers in the 6.7-GHz (5$_{1}$--6$_{0}$ A$^{+}$) transition \citep{Menten91} have been considered one of the best tracers of star formation. Methanol masers are divided into two classes, originally reflecting their different locations with respect to the associated young stellar object \citep{Menten91a}, but which are now understood to reflect their differing pumping mechanisms. The 6.7-GHz transition is the strongest and most widespread of all the class II (radiatively pumped) methanol masers and are considered to be exclusively associated with high-mass star formation regions \citep[e.g.][]{Minier03,Xu08,Gallaway13}. Although a number of observations have targeted low-mass star formation regions \citep[e.g.][]{GreenMMB11,Minier03,Bourke05} no sources of 6.7-GHz methanol maser emission have been detected to date, nor are they expected given the temperatures and methanol column densities required to produce class II methanol maser emission \citep[e.g.][]{Cragg05}. 

While it is widely accepted that low-mass young stellar objects (YSOs) are unable to produce class II methanol masers, the lower mass limit on the stars that can produce them is not well constrained.  \citet{Minier03} detected a low-luminosity 6.7-GHz methanol maser towards the Orion B region (NGC2024:FIR 4).  This is thought to be an intermediate mass protostar, although its true nature remains enigmatic (see discussion in \citeauthor{Minier03}).  High-resolution images of some 6.7-GHz methanol masers show simple velocity gradients which have been interpreted as being due to the molecular gas lying within a rotating disk \citep[e.g.][]{Minier98}.  The mass inferred when the linear extent and velocity gradient of the 6.7-GHz methanol masers is fitted by Keplerian rotation is typically less than 8 \Msol \citep[e.g.][]{Minier00,Goddi11}, the lower limit of what is generally considered a high-mass star.  However, this is likely the result of one or more of the many assumptions made in the fitting process (particularly that the masers trace the full extent of the disk), as the bolometric luminosity of the associated protostars and/or associated molecular outflows generally suggest the presence of a high-mass protostar at the location of the 6.7-GHz methanol masers.  In contrast to the class~II transitions, class~I methanol masers have been detected towards low-mass star formation regions \citep{Kalenskii10}, although they have much lower luminosities than those associated with high-mass star formation regions. Widespread class~I maser emission from the 36~GHz transition has also recently been detected towards the Galactic Centre region \citep{YZ13}.

From the observations outlined above we can clearly infer that class~I methanol maser transitions can be inverted in a variety of astrophysical environments, and are not exclusively associated with high-mass star formation.  Class~II methanol masers are radiatively pumped and generally have a much more restricted distribution within high-mass star formation regions than the class~I transitions \citep[e.g.][]{Cyg09}. Recently the exclusivity of the association of class II methanol masers with high-mass YSOs has been questioned by the suggestion that these masers could also be associated with evolved stars \citep{Urquhart13,Walsh03}. In both cases the absence of any detectable dust continuum emission is integral to the questions raised about the nature of the sources. 

\citet{Urquhart13} compared complete samples of methanol masers \citep{CasMMB10,GreenMMB10,CasMMB11} and 870-$\mu$m continuum emission \citep[ATLASGAL;][]{Contreras13,Schuller09}. They found 577 methanol masers had accompanying dust continuum emission and identified 43 methanol masers devoid of dust continuum emission. In the majority of cases \citet{Urquhart13} were able to identify diffuse, or weak compact emission at the location of the 43 methanol masers and concluded that the lack of an associated 870-$\mu$m point source \citep{Contreras13} was mostly due to these sources being located at much larger distances. \citet{Urquhart13} find that this explanation is unsatisfactory for a small number of sources, and remark that another possibility is that these masers arise in the circumstellar shells associated with evolved stars. 

Of the 43 methanol maser sources that \citet{Urquhart13} find to have no accompanying dust continuum emission \citep{Contreras13}, 10 had been reported previously at other wavelengths. One of these shows IRAS emission, two sources were detected in the Bolocam Galactic Plane Survey \citep[detected due to its sensitivity to low surface brightness objects being superior to ATLASGAL;][]{Aguirre11} and seven are listed in the intrinsically red source compilation of \citet{Robitaille08}. Of the seven sources in \citet{Robitaille08}, six are identified as YSOs and a further source is listed as a possible AGB star (G\,328.385+0.131). \citet{Urquhart13} use the point-like nature of the GLIMPSE 8.0-$\mu$m emission coincident with this source to assert that this object is indeed an evolved star, since they expect 8.0-$\mu$m emission associated with high-mass star formation regions to be extended.

Previously, \citet{Walsh03} conducted 450-$\mu$m and 850-$\mu$m dust continuum observations with SCUBA on the JCMT towards 71 methanol masers, identifying one 6.7-GHz methanol maser site that was not associated with any sub-mm continuum emission. The methanol maser, G\,10.10+0.73, was discovered in Parkes observations towards IRAS~18021--1950 \citep{Walsh97} and later followed up with the ATCA to determine a precise position \citep{Walsh98}. While \citet{Walsh03} failed to detect any sub-mm continuum emission associated with the methanol maser, they did detect a compact source within the target field (offset 90" from the methanol maser), associated with the bipolar planetary nebula NGC~6537. The authors question whether there is an association between the methanol maser and the planetary nebula, suggesting that the kinematics are well matched but maintaining that a chance alignment is a possibility. \citet{Urquhart13} claim that this source supports their inference that methanol masers may be associated with evolved stars.

Here we present some general arguments as to why evolved stars are unlikely to harbour 6.7-GHz methanol masers (Section~\ref{sect:meth}), briefly discuss the dust properties in star formation regions (Section~\ref{sect:dust}) and then focus our investigation on the two 6.7-GHz methanol masers, G\,328.385+0.131 (Section~\ref{sect:328}) and G\,10.10+0.73 (Section~\ref{sect:walsh}), where an association with evolved stars has been explicitly suggested \citep{Walsh03,Urquhart13}.

\section{Methanol and late-type stars}\label{sect:meth}

OH and water masers are observed in a wide range of astrophysical environments - from star forming regions (both low and high-mass in the case of water masers), evolved stars, supernovae remnants (1720-MHz OH) and the central regions of active galaxies.  The 6.7-GHz methanol transition has been shown to be strongly inverted over a broad range of physical conditions \citep[e.g.][]{Cragg+02}, similar to the ground-state OH and 22-GHz water transitions.  Therefore it is natural to ask the question: are there astrophysical environments, other than those associated with high-mass star formation regions, which can support 6.7-GHz methanol masers?

There are nearly 1000 known 6.7-GHz methanol maser sources \citep{CasMMB10,GreenMMB10,CasMMB11,GreenMMB11} and in excess of 2000 late-type stars known to exhibit maser emission - primarily found from searches for 1612-MHz OH \citep{Sevenster+97a,Sevenster97,Sevenster+01} or SiO \citep[e.g.][]{Deguchi07} maser transitions.  To date, no overlap between the 6.7-GHz methanol masers and evolved stars showing either 1612-MHz OH or SiO maser emission has been identified.  Given the large size of both maser samples, and their typical positional accuracies of $\sim$1 arc second, we can conclude that evolved stars showing 1612-MHz OH or 43-GHz SiO masers do not have conditions which are suitable for 6.7-GHz methanol masers.  

For a molecular transition to exhibit maser emission requires both suitable physical conditions (temperature, density, etc), and also a sufficient abundance of the relevant molecule.  As evolved stars host maser emission in a range of ground-state OH transitions, as well as water and SiO, it would seem likely that some regions within that environment will have the temperatures and densities  within the broad range necessary to support 6.7-GHz methanol masers.  The critical difference between methanol and either OH or water seems to be that methanol is less readily produced in high abundance than either OH or water.

\citet{Garrod+06} summarises the evidence as to why methanol is thought not to be produced in gas phase reactions in the interstellar environment.  The favoured mechanism for the formation of methanol is through hydrogenation of CO when it freezes out on dust grains \citep[e.g.][]{TW97}.  Low grain temperatures favour the formation of formaldehyde and methanol rather than CO$_2$ \citep{HM99,Taquet+13}, and in some very young star formation regions infrared spectroscopy has found methanol ice to be the second most abundant ice on dust grains, after water \citep{Dartois+99}.  Methanol ice has been observed towards both low- and high-mass star formation regions \citep[e.g.][]{Pontoppidan+04,Gibb+04}.  As the forming stars heat the surrounding medium the methanol desorbs from the dust grains and is released into the gas phase. Mechanical desorption of the dust grains by outflows is also possible \citep[e.g.][]{Gibb98} and can play an important role in increasing the methanol abundance in the gas phase in some sources.

Similar to star formation regions, the environment near evolved low-mass stars contains large amounts of dust, however, it is not expected to be at temperatures of around 10~K, which favour the formation of methanol on the grain mantles \citep[e.g.][]{Taquet+13}. Amongst evolved-star maser sources, a small number show emission in atypical maser transitions, for example the proto-planetary nebulae K 3-35 possesses 1720 MHz and 6 GHz excited OH masers \citep{Desmurs10}.  Some very young planetary nebulae and proto-planetary nebulae also have associated cold dust \citep{Sahai97}, however, the dust masses are very much less than 1~\Msol\ and are unlikely to facilitate the formation of large amounts of methanol ice. To date there have been no detections of methanol towards any evolved stars, despite sensitive measurements towards a number of nearby sources \citep[e.g.][]{Ford+04,He+08}.  In particular, observations towards the extreme carbon star IRC+10216 (towards which more than 50 molecular species have been observed), failed to detected any methanol with a 3-$\sigma$ abundance upper limit of $8.5 \times 10^{-10}$ (although the same observations detected formaldehyde with more than an order of magnitude greater abundance).  

In summary, current understanding of the formation process for methanol in astrophysical environments suggest that it is unlikely to be present in the vicinity of evolved stars and this is supported by existing observational studies.

\section{Dust emission from young stars}\label{sect:dust}

\citet{Hill05} made 1.2-mm dust continuum observations of a sample of 131 high-mass star formation regions, identified on the basis of either the presence of an \UCHII region, or a 6.7-GHz methanol maser. Similar to \citeauthor{Urquhart13}, \citeauthor{Hill05} found a small fraction of their targets (20 methanol masers and 9 \UCHII regions) for which they did not detect dust continuum emission.  At the 3-$\sigma$ sensitivity of their observations, \citeauthor{Hill05} were sensitive to all sources with a dust mass in excess of 600~\Msol\ at distances of 16.3~kpc or less (i.e. over the vast majority of the region of the Galaxy where high-mass star formation is present). Since the majority of the dust sources they detected have masses in excess of their sensitivity limit, they suggest three possible explanations for the 29 non-detections: 1) these sources have characteristics dissimilar to the majority of high-mass star formation regions; 2) that these sources are associated with later stages of star formation; or 3) these sources are located at too great a distance and are not massive enough to be detectable.

At least two of the nine \UCHII regions for which \citet{Hill05} find no accompanying dust continuum emission have been extensively studied \citep[G\,188.770+1.074 and G\,23.43--0.18 e.g.;][]{Kurtz94,Walsh98,KK01,Trinidad10} and their \UCHII region nature is irrefutable. Both of these sources are located in complexes which contain methanol masers with distances measured through trigonometric parallax observations \citep[G\,188.95+0.89 and G\,23.44--0.18;][]{Reid09}. G\,188.770+1.074 and G\,23.43--0.18 are located at 2.10 and 5.88~kpc \citep{Reid09}, respectively, meaning that the \citet{Hill05} observations are not sensitivity limited. This clearly demonstrates that some high-mass star formation regions have unusually low dust masses, by the evolutionary phase when \UCHII regions and 6.7-GHz methanol masers are present. Thus, the absence of dust emission associated with a methanol maser is not itself sufficient to suggest an origin other than a high-mass star formation region.

\section{G\,328.385+0.131 and G\,10.10+0.73 - associated with evolved stars?}\label{sect:ATLASGAL}

We have investigated the two maser sources for which an association with an evolved star has been proposed \citep{Urquhart13,Walsh03}. The results of our investigation of G\,328.385+0.131 are given in detail in Section~\ref{sect:328}. The case of G\,10.10+0.73 can be simply dismissed as a spurious detection as discussed in Section~\ref{sect:walsh}. Here we summarise the key arguments which clearly show that there is no basis to the proposition that G\,328.385+0.131 is associated with an evolved star.

\begin{itemize}
\item The classification of G\,328.385+0.131 by \citet{Robitaille08} as a possible AGB star is likely erroneous. The source is most probably a luminous YSO which their mid-infrared colour-magnitude criteria misclassified. 
\item The infrared properties of G\,328.385+0.131 are in the middle of the range seen for large samples of methanol masers. We further find the point-like nature of the associated 8.0-$\mu$m emission to be fairly common for sources associated with methanol masers.
\item The methanol maser at G\,328.385+0.131 has a high luminosity. If such a luminous maser were associated with an evolved star we would expect to see other examples within the Galaxy.
\item We find a convincing $\sim$3-$\sigma$ 870-$\mu$m detection ($\sim$320~mJy) at G\,328.385+0.131 in the ATLASGAL data, contrary to the non-detection reported by \citet{Urquhart13}, which surpasses the flux density possible from any AGB star. 
\item The mass of the dust associated with G\,328.385+0.131 is $\sim$600~M$_{\odot}$ which is consistent with other regions that support high-mass star formation.
\end{itemize}

\subsection{Methanol maser G\,328.385+0.131}\label{sect:328}

\subsubsection{Infrared characteristics}

The possible AGB star designation assigned to G\,328.385+0.131 by \citet{Robitaille08} was determined from a set of criteria they derived for the separation of YSOs and AGB stars. 
\citet{Robitaille08} classify sources with [4.5] $\leq$ 7.8 as `extreme' AGB star candidates, but note that this criterion will also capture a number of luminous YSOs. In the case where [4.5] $>$ 7.8 there are two further criteria: if [8.0] -- [24.0] $<$ 2.5 they classify the source as a `standard' AGB star, and if [8.0] -- [24.0] $\geq$ 2.5 they classify the source as a YSO. Examining the relevant colours and magnitudes of G\,328.385+0.131, it is clear that it barely makes it into the possible AGB category on the first criterion, with [4.5] = 7.68 (in fact, this value is only 3.2-$\sigma$ from the 4.8 micron magnitude threshold of 7.8). If this source had a slightly higher magnitude, it would be classified as a YSO as it satisfies the second criterion ([8.0] -- [24.0] = 3.5). It is therefore likely this source is one of the luminous YSOs that has been wrongly classified as an AGB star, as \citet{Robitaille08} mentioned was inevitable. 

\citet{Gallaway13} used an adaptive non-circular aperture photometry technique to determine the GLIMPSE fluxes for sources associated with the methanol masers detected in the Methanol Multibeam Survey \citep[a survey for methanol maser emission in the Galactic Plane;][]{Green09}. Their table~1 lists 4.5-$\mu$m magnitudes towards 444 methanol masers, a large fraction of which (35 per cent) have [4.5] $\leq$ 7.8 and would therefore also be classified as `extreme' AGB stars by the \citet{Robitaille08} criteria. It is clearly unreasonable to expect such a large fraction of methanol masers to be associated with AGB stars, when an alternative explanation exists in which these are instead the expected fraction of YSOs that are inevitably captured by the simplistic colour-magnitude criteria \citep{Robitaille08}. This highlights the difficulties in relying upon infrared selection criteria alone to classify individual sources. Furthermore, looking more closely at the IRAC colours of the mid-infrared counterpart of G\,328.385+0.131, it is evident that this source lies in the middle of the range of colour-colour and colour-magnitude plots of sources associated with methanol masers \citep{Gallaway13,Breen11,Ellingsen06} adding further support to the argument that this source is actually a standard high-mass star formation region.

In their study of the GLIMPSE properties of sources associated with 6.7-GHz methanol masers, \citet{Gallaway13} also found that of their 769 methanol masers, 219 (28 per cent) were well characterised by a point source in all four IRAC bands. Since this is the case, the point-like nature of the infrared source associated with G\,328.385+0.131does not indicate that the emission is associated with an AGB star, as suggested by \citet{Urquhart13}.

\subsubsection{Distance, maser luminosity and association with other lines}

Our ability to confidently estimate the distance of G\,328.385+0.131 is dependent on the nature of the source: if it is a high-mass star formation region it is necessarily constrained to the spiral arms and the velocity of the methanol maser emission will follow that of Galactic rotation \citep[6.7-GHz methanol masers typically show their median velocity within 3 -- 5~\kmsns of the systemic velocity of the region;][]{Szy07,Pandian09,GM11}, making kinematic distance determination viable; if, on the other hand, the source is an AGB star, the range of expected velocities is not restricted to Galactic rotation and kinematic distance determination is subject to much greater uncertainty.

\begin{figure}
	\psfig{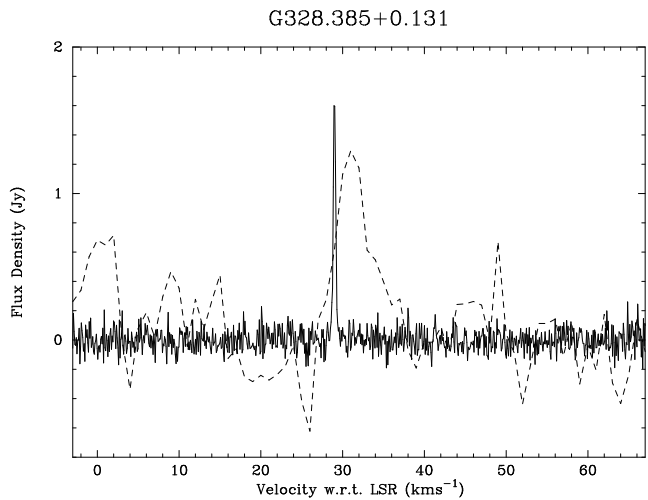}
	\caption{The methanol maser G\,328.385+0.131 \citep{GreenMMB11} overlaid with the CO emission (a 6.8-$\sigma$ detection) from the nearest pixel of the NANTEN data (position G\,328.400+0.133). Note that the CO spectrum has the units of K, but the same scale as the methanol maser.}
\label{fig:co_meth}
\end{figure}

We have inspected the CO emission from the NANTEN Galactic Plane Survey \citep{Mizuno04} and find an isolated spectral feature in the CO spectrum that matches the velocity of the methanol maser emission (see Fig.~\ref{fig:co_meth}). The spectral structure of the detected CO emission is unlike typical AGB CO spectra \citep[e.g.][]{CC+10} which tend to be broad (FWHM $>$20~\kmsns), flat-topped and steep-edged (i.e. more like a top-hat than a Gaussian profile). Crucially, if the CO was associated with an AGB of an anomalous velocity we would expect to see the CO emission restricted only to the immediate spatial vicinity of the AGB star. \citet{CC+10} measured the CO properties of a representative sample of AGB and post-AGB stars, finding that the (1-0) emission was extended by no more than $\sim$40 arcsec for sources at distances up to 1~kpc. The NANTEN survey was observed in a 4 arcmin grid and the CO emission presented in Fig.~\ref{fig:co_meth} extends to at least two positions adjacent to G\,328.400+0.133, corresponding to a structure of CO covering at least 8 arcmin. Such a large structure is wholly inconsistent with the CO emission expected for any AGB star, even one that is very close by. Conversely, such a CO structure is completely consistent with the extended emission seen within the spiral arms of our Galaxy, further confirmed by the velocity of the emission, which matches that of the Carina-Sagittarius spiral arm at this Galactic longitude. 

Given the tendency of methanol masers to trace the systemic velocities of the regions they are associated with, the well matched velocities of the methanol and CO emission, together with the expected velocity of the Carina-Sagittarius spiral arm at this longitude and the low latitude of the source, it is most likely that the methanol maser is associated with CO following Galactic rotation. Thus the positive velocity of the methanol maser at 28.9~\kms \citep{GreenMMB11} undoubtedly places it at the far kinematic distance, in agreement with the distance assigned by \citeauthor{Urquhart13} of 18.0~kpc. We further note the presence of nearby luminous SNR 328.4+0.2, which is located at a similar distance (16.8~kpc: \citet{Caswell75}; 17.4$\pm$0.9~kpc: \citet{Gaensler00}), illustrating that this part of the Galaxy is clearly capable of supporting high-mass star formation; in fact, the Galactic longitude versus velocity diagram presented in \citet{GreenMMB11} shows that there are several other methanol masers with similar Galactic longitudes and velocities. 

The detected CO emission implies a H$_2$ mass of $\sim$5300~\Msol\ \citep[assuming a solar-neighbourhood Galactic X-factor of 1.8$\times$10$^{20}$;][and a distance of 18~kpc]{Dame+01} in a single NANTEN position, or a total cloud mass of $\sim$10$^4$~\Msol\ (over the three adjacent positions where CO at this velocity was detected). The estimated mass of the molecular gas is also consistent with the cloud being located within the spiral arms of our Galaxy.

The association of the methanol maser with the co-rotating gas within the spiral arms of our Galaxy does not alone definitively suggest either an AGB or high-mass star formation origin. However, the strong likelihood of a far kinematic distance ($\sim$18~kpc) is key to arguments made later in this section and those in Section~\ref{sect:dust}.

%
%
%

The methanol maser emission associated with G\,328.385+0.131 was reported by \citet{GreenMMB11} to consist of a single spectral feature with a peak flux density of 1.6~Jy, and appeared stable over the course of the initial MMB survey and followup observations. While this is a relatively modest flux density, its distance at 18~kpc places it in the upper middle of the luminosity range of all 366 MMB sources presented in fig. 7 of \citet{Breen11}. If an AGB star were capable of producing such a luminous methanol maser, we would expect to detect further sources of this type within the Galaxy, but we do not. Comparison between a large fraction of the full complement of 6.7-GHz methanol masers in the Southern hemisphere and mid-infrared data lead \citet{Gallaway13} to reconfirm that these masers are exclusively associated with high-mass star formation. If there were a significant population of methanol masers associated with evolved stars, it would be expected that the \citet{Gallaway13} study (and other earlier studies) would have uncovered it.

No further indications of the source's nature can be derived from the targeted 12.2-GHz methanol maser observations conducted by \citet{Breen12}, which revealed no emission at 5-$\sigma$ levels of 0.75 and 0.8~Jy during 2008 June and December, respectively. The 12.2-GHz methanol maser catalogues of \citet{Breen12a,Breen12} show that in the Galactic longitude range 186 to 330$^{\circ}$ approximately 44 per cent of 6.7-GHz methanol masers have 12.2-GHz counterparts (commonly with 6.7-GHz to 12.2-GHz peak flux density ratios greater than 2:1). 

We have inspected the water maser \citep{Walsh11} and ammonia data \citep{Purcell12} from the H$_2$O Southern Galactic Plane Survey (HOPS) and find no detection of either tracer. The modest sensitivity of this survey (98 per cent complete at 8.4~Jy for water masers and 100 per cent complete at $T_{mb}$ = 1.0~K for ammonia) combined with the outer Galactic distance for this source preclude us from drawing any conclusions from these data. Recent observations for OH masers in the 1612-, 1665-, 1667- and 1720-MHz lines have been carried out as part of MAGMO \citep[a project to study the Magnetic fields of the Milky Way through OH masers;][]{Green12} towards this source and made no detections in any of the OH transitions to a 5-$\sigma$ detection limit of $\sim$250~mJy.

\subsubsection{The dust emission}\label{sect:dust}

Using ATLASGAL data, \citet{Urquhart13} report that there is no 870-$\mu$m emission at the location of the methanol maser to an upper limit of 320~mJy. We have inspected the image provided by \citeauthor{Urquhart13}, which shows the mid-infrared environment associated with G\,328.385+0.131 overlaid with ATLASGAL contours. This image shows unresolved emission at a 3-$\sigma$ level (signified by the presence of three 1-$\sigma$ interval contours) towards the location of the methanol maser. Given the coincidence of the dust emission with the methanol maser, and the absence of any further emission beyond a 1-$\sigma$ level in the region surrounding the methanol maser, we believe this to be an authentic dust continuum detection. We therefore reinterpret the \citeauthor{Urquhart13} upper limit of 320~mJy to be a detection at about this flux density and similarly the presented dust mass upper limit of 579.17~\Msol\/beam$^{-1}$ to be the estimated dust mass of the detected source.

 \citeauthor{Urquhart13} find that 28 per cent of dust sources with methanol masers have calculated dust masses of less than 1000~\Msol. They suggest that the majority of these are more compact (in most cases $<$0.3~pc) and may have higher star-formation-efficiency and be forming smaller-stellar-mass systems. From figure 15 of \citeauthor{Urquhart13}, it seems that all of their sources with masses greater than about 100~M$_{\odot}$ fall in the high-mass star forming range of values satisfying the \citet{Kauf10a,Kauf10b} criteria. There is therefore no evidence to suggest that G\,328.385+0.131, with a mass of $\sim$550~\Msol\/beam$^{-1}$, would be unable to support high-mass star formation. Furthermore, \citeauthor{Urquhart13} apply a constant temperature of 20~K to all of their sources. While this is a reasonable average for many young star formation regions, individual sources may vary by several K from this value which significantly impacts the derived mass (e.g. if the temperature was actually 15~K, as can often be the case \citep[e.g.][]{Carey00}, the mass would increase by 58 per cent).

Conversely, the likelihood that an AGB star would be detectable at such a far distance in the 870-$\mu$m ATLASGAL data seems unlikely. To consider if this is possible, we have calculated the expected 870-$\mu$m emission from an AGB star located at 18~kpc. The flux density of the dust envelope of an AGB star at 870-$\mu$m, assuming that the dust continuum emission is optically thin, is given by \citep{Hildebrand83}:
\begin{equation*}
F_{870}=1.05\times10^{36}\frac{M_d \kappa_{870} B_{870}(T_d)}{D^2},
\end{equation*}
\noindent where $M_d$ is the AGB envelope mass, $\kappa_{870}$ is the dust opacity coefficient, $B_{870}(T_d)$ is the Planck function at the dust temperature (ergs s$^{-1}$ cm$^{-3}$), and D is the distance (pc). 

The dust opacity at 870-$\mu$m has an estimated value of 1.87~cm$^{2}$~g$^{-1}$ for ATLASGAL observations \citep{Schuller09}, however, toward circumstellar grains it can vary by up to an order of magnitude \citep{Sopka85}. \citet{Ladjal10} conducted 870-$\mu$m observations towards nine evolved stars and found values of dust opacity ranging from 2 to 35 cm$^{2}$ g$^{-1}$. We therefore calculate limits on the expected flux density using these two extreme values.

Assuming a dust temperature of 1000~K  \citep[estimated from observations of nearby AGB stars;][]{Ladjal10}, together with their average dust envelope mass of $\sim$2$\times$10$^{-3}$ M$_{\odot}$, the expected flux densities for each of the two extreme dust opacity values are:

\begin{equation*}
F_{870} =\begin{cases} 9~$mJy$~($for$~\kappa_{870} = 2~$cm$^{2}~$g$^{-1})\\
160~$mJy$~($for$~\kappa_{870} = 35~$cm$^{2}~$g$^{-1}) \end{cases},
\end{equation*}

\noindent making it is very unlikely that 870-$\mu$m emission would be any stronger than 160~mJy and quite likely considerably less given a more reasonable value of dust opacity. Since the flux density of the 3-$\sigma$ 870-$\mu$m emission associated with the methanol maser is $\sim$320~mJy, it is highly probable that we can rule out that the origin of this emission is an AGB star.

%
%

\subsubsection{SED modelling}

\begin{figure}
	\psfig{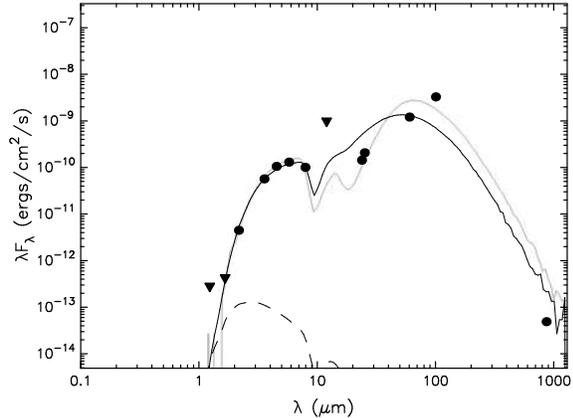}
	\caption{The SED and fitted YSO models for G\,328.385+0.131. The dots and the triangles are measurements and upper limits taken from (left to right) the 2MASS (3), GLIMPSE (4), IRAS (1), MIPSGAL \citep[1;][]{Robitaille08}, IRAS (3) and ATLASGAL (1) surveys. The solid black line shows the best fitting model, the solid grey line gives the second best fitting model, while the dashed line gives the spectrum of the stellar photosphere in the absence of circumstellar dust but accounting for interstellar extinction. }
\label{fig:sed}
\end{figure}

We have used the online SED fitting tool of \citet{Robitaille+07} to investigate the nature of the emission associated with G\,328.385+0.131. We allowed the fitting distance to vary between 16 and 18 kpc. The resultant SED fit is presented in Fig.~\ref{fig:sed}, and shows the characteristics of a deeply embedded, high luminosity YSO. Using the criterion of \citet{Carlson+11}, we were able to confirm that there was no significant PAH emission, which is not accounted for in the YSO models used in the SED fitting \citep{Robitaille+06}.

\citet{Pandian10} investigated the SEDs of the centimeter through to near-infrared emission associated with 20 6.7-GHz methanol masers which comprise a complete sample of methanol masers between Galactic longitudes 38.6$^{\circ}$ and 43.1$^{\circ}$, and latitudes $|$b$|$ $\leq$ 0.42$^{\circ}$. The fitted characteristics of G\,328.385+0.131 (derived from the best fitting model shown in Fig.~\ref{fig:sed}) all fall within the range of those derived for the \citeauthor{Pandian10} sample. The only exception is the source age which is about twice that of the oldest sources given in their study. The significance of this difference is unlikely to be high given the size of their sample, together with the range of their values (1.2 to 57~$\times$~10$^3$ years c.f. 110~$\times$~10$^3$ years for G\,328.385+0.131).

\citet{Robitaille08} present a sample of SEDs for both their standard AGB and YSO candidates. Although some of the `extreme' AGB candidates presented in \citet{Robitaille08} have SEDs that are more comparable to that of YSOs, the standard AGB candidates have very different characteristics than those of G\,328.385+0.131 shown in Fig.~\ref{fig:sed}. In each of the 18 standard AGB cases presented in \citet{Robitaille08}, the SED falls significantly beyond wavelengths of 5.8-$\mu$m (and sometimes from even shorter wavelengths), a property not shared by the data presented for G\,328.385+0.131 which is well fitted by a YSO model.

\subsection{Methanol maser G\,10.10+0.73}\label{sect:walsh}

\citet{Walsh98} conducted ATCA observations of methanol masers detected towards IRAS sources \citep{Walsh97} in order to derive precise source positions. The methanol maser observed by \citet{Walsh98} in the target field of IRAS source 18021--1950 was found to be significantly offset from the target IRAS source (by 116.79" in RA and --38.46" in declination) and was detected as a single feature of 1.2~Jy at a velocity of 1.2~\kmsns. The location of this methanol maser was found to have no coincident dust continuum emission \citep{Walsh03} and, combined with its proximity to planetary nebula NGC~6537 led to the suggestion of a possible association between the two objects. 

This methanol maser was not detected in the complete MMB survey \citep{GreenMMB10}, perhaps not too surprising given the intrinsically variable nature of masers and the 5-$\sigma$ sensitivity of the survey observations, which at 0.7~Jy is not greatly different from the flux density of the \citet{Walsh98} detection. Although the non-detection in the \citet{GreenMMB10} observation is not suspicious alone, the reported maser is, notably, less than a degree from the strongest methanol maser in the Galaxy, G\,9.621+0.196, with velocity of 1.3~\kms and emission regularly detected with a peak flux density in excess of 5000~Jy \citep[e.g.][]{GreenMMB10}, and this prompted us to download and inspect the data used in \citet{Walsh98} from the Australia Telescope Online Archive (ATOA).

The ATCA data was taken on 1994 July 30 and consisted of four cuts of two minutes each, spread over almost 8 hours. Observations of PKS~B1934--638 were made for primary flux density and bandpass calibration, and 1808--209 for gain calibration (which had a flux density of $\sim$0.25~Jy at the time of the observations). For these observations the correlator was set to record 1024 channels over a bandwidth of 4-MHz, yielding a channel spacing of 0.18~\kms and a coverage of 180~\kms \citep{Walsh98}.

Despite several attempts using {\sc miriad}, employing standard data reduction techniques for ATCA spectral line data, we have been unable to reproduce the 1.2~Jy detection from the same ATCA data. We suggest that the reported detection is most likely an image artefact arising from the nearby bright of G\,9.621+0.196 which was more pronounced than usual because the quality of the data was inadequate to form reliable images (in particular, the phase calibrator was too weak for the correlator configuration and system available at the time of the observations) and possibly further due to inferior data reduction algorithms available almost 20 years ago. 

We conclude that the reported methanol maser G\,10.10+0.73 detected at a velocity of 1.2~\kms is not an authentic maser detection and therefore has no possibility of supporting the suggestion that methanol masers can be associated with evolved stars as suggested in both \citet{Walsh03} and \citet{Urquhart13}.

\section{Conclusion}

We have investigated the suspected association between two 6.7-GHz methanol masers, G\,328.385+0.131 and G\,10.10+0.73, with evolved stars \citep{Urquhart13,Walsh03}. We find no evidence to support these claims, rather finding G\,328.385+0.131 to be an average methanol-maser-associated YSO, and  G\,10.10+0.73 to be a spurious source caused by the strong emission from nearby methanol maser G\,9.621+0.196. 


More generally, we argue that it is unlikely that the environments of evolved stars contain a high enough abundance of methanol to support maser emission. We therefore confirm that the exclusive association of 6.7-GHz methanol masers with star formation regions remains valid.

\section*{Acknowledgments}

We thank the referee, Anita Richards, for insightful and thorough comments that allowed us to significantly improve the quality of the paper. This research has made use of: NASA's Astrophysics
Data System Abstract Service; the NASA/
IPAC Infrared Science Archive (which is operated by the Jet Propulsion
Laboratory, California Institute of Technology, under contract with
the National Aeronautics and Space Administration); the SIMBAD data base, operated at CDS, Strasbourg,
France; and data products from the GLIMPSE and MIPSGAL
surveys, which are legacy science programs of the {\em Spitzer Space
  Telescope}, funded by the National Aeronautics and Space
Administration.

\end{document}